**AI analysis of super-resolution microscopy: Biological discovery in the absence of ground truth**

Ivan R. Nabi[1,2*], Ben Cardoen[3], Ismail M. Khater[3,4], Guang Gao[1], Timothy H. Wong[1], Ghassan Hamarneh[3*]

[1]Department of Cellular & Physiological Sciences, Life Sciences Institute, University of British Columbia, Vancouver, BC, Canada V6T 1Z3
[2]School of Biomedical Engineering, University of British Columbia, Vancouver, BC, Canada V6T 1Z3
[3]School of Computing Science, Simon Fraser University, Burnaby, BC, Canada V5A 1S6
[4]Department of Electrical and Computer Engineering, Faculty of Engineering and Technology, Birzeit University, Birzeit P627, Palestine

*to whom correspondence should be addressed: irnabi@mail.ubc.ca; hamarneh@sfu.ca .

Short title: AI, nanoscopy and biological discovery

**Abstract**

Super-resolution microscopy, or nanoscopy, enables the use of fluorescent-based molecular localization tools to study molecular structure at the nanoscale level in the intact cell, bridging the mesoscale gap to classical structural biology methodologies. Analysis of super-resolution data by artificial intelligence (AI), such as machine learning, offers tremendous potential for discovery of new biology, that, by definition, is not known and lacks ground truth. Herein, we describe the application of weakly supervised paradigms to super-resolution microscopy and its potential to enable the accelerated exploration of the nanoscale architecture of subcellular macromolecules and organelles.

**eTOC Blurb**

Application of artificial intelligence (AI) to super-resolution microscopy accelerates discovery of novel biology that necessarily lacks ground truth. Herein, we discuss supervision paradigms for biological discovery from super-resolution microscopy to enable AI-accelerated exploration of the nanoscale architecture of subcellular macromolecules and organelles.

Artificial intelligence (AI), “the capability of computer systems or algorithms to imitate intelligent human behavior” (Merriam-Webster.com, 2024) is increasingly present in our everyday lives. Recently, generative language and image AI models (ChatGPT, Sable Diffusion, Midjourney) created a storm of interest, with users challenged to differentiate between AI-generated and human-generated content. Users are both amazed by AI yet disappointed by its occasional surprising failures and hallucinations. Like ChatGPT, current AI methods are incredible tools that are not always correct, necessitating expert “ground truth” validation.

“Ground truth is information that is known to be real or true, provided by direct observation and measurement (i.e., empirical evidence) as opposed to information provided by inference” (Wikipedia, 2024). For imaging-based AI applications, validation uses ground truth image annotations to test computer identification of images and image components, from distinguishing cats from dogs to real-world examples, such as medical imaging-based diagnosis or object recognition for self-driving cars. Machine learning models are trained and tested on acquired datasets. Supervised machine learning uses ground truth annotations to train new image prediction methods and validate predictions.

The most straightforward approach to train machine learning to identify objects within an image is to perform strong supervision. The machine learning model is trained on a curated dataset of images and their corresponding segmentation masks (dense, pixel/voxel-level annotations); class labels are assigned to every pixel in the image. However, when applying AI to novel bioimaging modalities, such as super-resolution microscopy (SRM) that breaks Abbe’s diffraction limit, even experts are challenged to define what is real within these images. As opposed to labeling street signs, vehicles, pedestrians, etc. for self-driving cars, for which we are all experts in principle, the time and financial cost of having expert biologists annotate images, at the pixel-, voxel-, or localization-level, can be astronomical. For SRM, strong supervision based on complete annotation is rarely feasible, with the noted exception of simulated data or phantom data such as DNA origami. While application of machine learning to SRM has tremendous potential to address unanswered questions and discover novel biology, ground truth expert annotation of image content is often infeasible. Annotation also relies on the assumption that experts know all there is to know about the underlying biology that these images capture, an assumption that may not always hold true. This is particularly the case for novel imaging modalities (such as SRM) whose primary purpose is to expand the boundaries of our understanding of biology (Figure 1 A).

The `strength’ of supervision describes the detail of knowledge that is known about and provided with the data to provide the training corpus. For example, in the case of classifying objects in an image, absence of supervision would mean there is no information provided about the class of each image, or about the objects depicted by the image. Weak supervision would describe a situation where we present information that, say, a dog or cat, is present in an image, but not where. Strong supervision would be a case where the machine learning method is provided with the knowledge of the exact location and outline of each identified object. When the supervisory signal is created from the input image itself without any annotation burden, we have self-supervision. For example, providing rotated images along with an automatically generated annotation ‘angle of rotation’ per image, or providing an image with added noise to it, along with the known original image, allows the machine learning method to learn how to cancel the rotation or remove the noise. By doing so it is possible to train a model to encode semantics of the objects in the images and leverage such knowledge to other tasks. While some fine tuning is invariably needed, a key advantage of self-supervised learning is that far less strongly supervised data is required for equal or better performance. The strength of supervision spans a spectrum, e.g., providing contours around each object is stronger than providing bounding boxes, which, in turn, is stronger than providing only the count of objects. It is also possible to mix different forms of supervision, e.g., a machine learning method may leverage self-supervision, as well as some images with a variety of weak annotations, and other images with strong supervision. Unsupervised methods do not use any annotations (e.g., clustering). Semi-supervised methods combine supervised and unsupervised datasets to train machine learning models (Figure 1 B).

In contrast to the high-annotation-burden of strong supervision, weak supervision assigns a single class label to the whole image rather than to every pixel (i.e. identifying a dog without specifying which pixels are part of the dog). An example of weak supervision in SRM is the provision of a training set of SRM images, along with a cell group or condition acting as the image class label. Dictating the image-level label does not involve annotating which pixels of the image are manifestations of that cell group or condition. Weakly supervised object detection and localization (i.e., training AI to find the object-specific locations by training on data labeled at the image level) has been popular for natural images (Zhou, 2017), and applied to biomedical images, such as MRI, histopathology and confocal microscopy (D'Alonzo et al., 2021; Liu et al., 2022a; Xu et al., 2014). We suggest that this form of supervision is suited to SRM, given that the goal is to identify and characterize those subcellular structures that vary across experimental conditions (cell lines, gene expression, mutations, infection, and drug treatment).

For biological research exploring the subcellular space, including macromolecules, organelles and cytoskeletal structures, ground truth has long been defined by high resolution approaches, in particular electron microscopy (EM). EM provides exceptional resolution, less than 1 nm, and pioneering EM work from the 1950’s revolutionized our understanding of cellular organelles, defining the morphological underpinnings of the cell that are now textbook cell biology. EM has provided unprecedented 3D views of the cell, such that we now understand that the cytoplasm is a complex and dense array of membrane-bound and non-membrane bound organelles organized amongst cytoskeletal elements. Indeed, Golgi’s discovery of the Golgi apparatus in neurons of silver-stained nervous tissue in the late 1800’s was challenged as an artifact until it was confirmed by EM in the 1950’s (Bentivoglio and Mazzarello, 1998). This represents perhaps the first example of EM being used as ground truth validation.

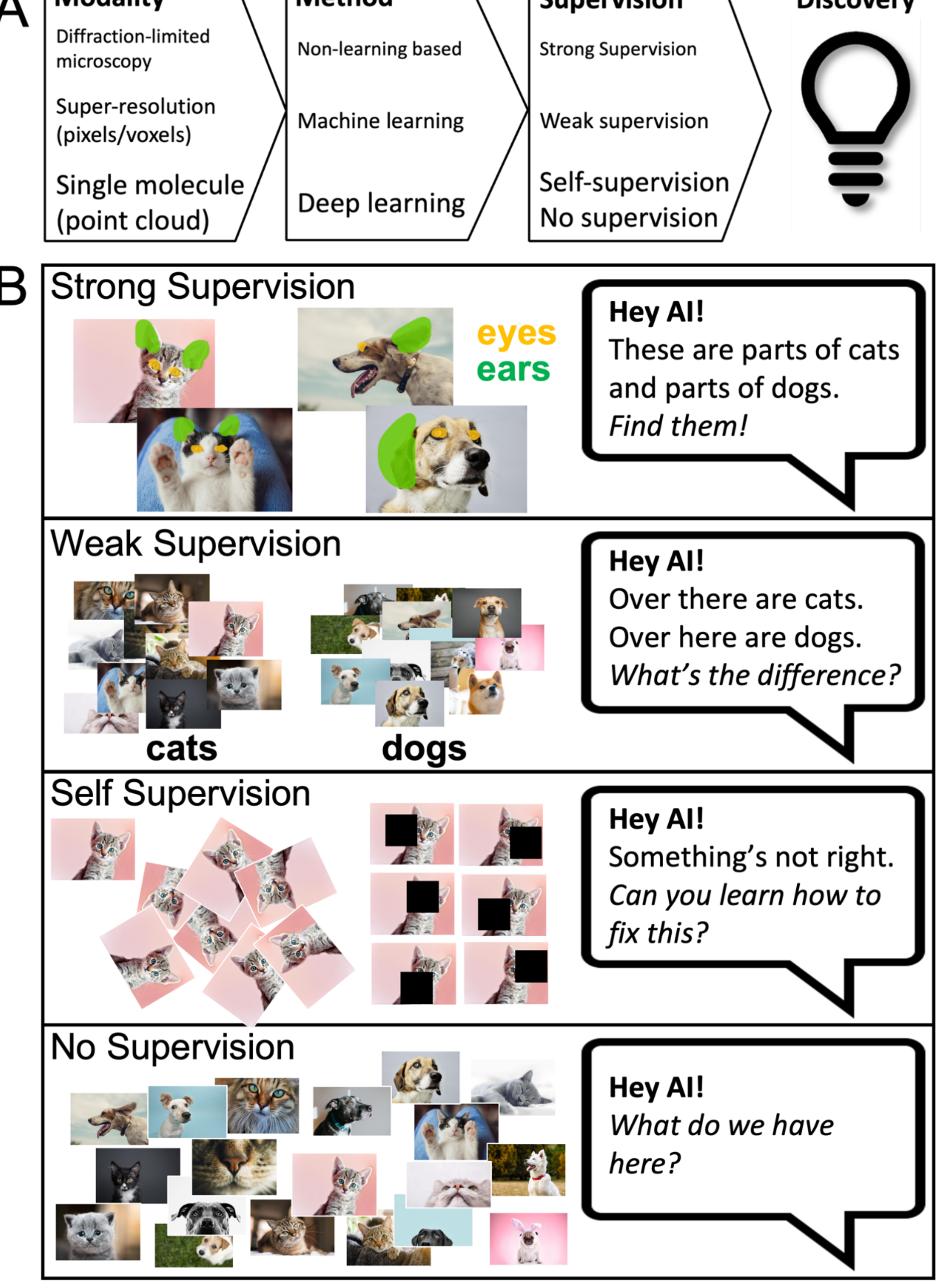

***Figure 1. AI and SRM.*** *A. Continued and parallel improvement in SRM hardware and AI-based analysis will lead to new and better approaches to define the organization and dynamics of sub-cellular structures. For instance, analysis of higher resolution single molecule approaches, such as MinFlux (2 nm lateral resolution) (Balzarotti et al., 2017) with deep learning and self-supervision modalities should lead to novel*

*insights into molecular structure in whole cell analyses. B. Different levels of supervision include strong supervision in which each object or even each pixel has annotated information that can be leveraged. This is rare in SRM as discovery implies absence of such data. Weak supervision occurs when partial information, often at the image level, is available, such as the presence or absence of a type of object, but not its location. In SRM use cases, this would be cell line, treatment, gene expression etc. Self-supervision is a hybrid form, where the model learns to operate on images using indirect information, such as rotations. This is then followed by a fine-tuning stage with small amounts of strongly supervised data. Unsupervised, i.e. no supervision, indicates a complete absence of such metadata.*

***Structural biology and super-resolution microscopy: closing the mesoscale gap***

Advances in structural biology technologies provide powerful tools to decipher molecular structures at the atomic level. X-ray crystallography and nuclear magnetic resonance (NMR) spectroscopy generated a wealth of molecular structural data that propelled the growth of the Protein Data Bank in the past 40 years. This enabled the development of co-evolution and AI-based algorithms such as AlphaFold2 and RoseTTAFold that are capable of accurately predicting structures of proteins from amino acid sequence alone for many targets (Jumper et al., 2021). The recent resolution revolution in cryo-EM made it possible to use the single-particle approach or subtomogram-averaging approach to determine, at atomic resolution, the structures of a broad range of macromolecular complexes from ribosomes to intact virions, filling the gap between sub-nanoscale to the mesoscale (Ke et al., 2020; Rozov et al., 2019; Zhang et al., 2010). However, with the exception of subtomogram-averaging, these conventional structural biology approaches are primarily used for characterizing highly purified samples that have been removed from their native subcellular or cellular environment (Figure 2). Further, EM is limited to fixed cell analysis, requires extensive sample preparation and is time-intensive. As such, while EM can provide ground truth validation, it cannot, however, provide the ground truth annotation required for strongly supervised learning training in super-resolution.

More rapid image acquisition, improved antibody labeling efficiency and availability of fluorescent proteins for live cell analysis make fluorescent microscopy the method of choice for analysing molecular distribution and dynamics in whole cell 3D volumes (Lippincott-Schwartz, 2011). Confocal microscopy improved axial resolution, to about 500 nm, and facilitated the use of dynamic photobleaching assays (i.e. fluorescence recovery after photobleaching (FRAP), and fluorescence loss in photobleaching (FLIP)) making it a routine tool to study subcellular structure and dynamics in many labs (Diekmann and Hoischen, 2014). While improving axial resolution, confocal microscopy does not address the diffraction barrier that limits lateral resolution of fluorescence microscopy to about 200-250 nm. SRM, defined as microscopy approaches that break the diffraction limit of light, encompasses a number of distinct methodologies enabling nanoscale fluorescent microscopy and provided novel insight into subcellular structure and dynamics (Sahl et al., 2017; Sydor et al., 2015). With cryoEM now able to analyze whole viruses at sub-nanometre resolution (Zhang et al., 2010) and SRM approaches such as MinFlux able to image down to 2 nm resolution (Balzarotti et al., 2017), the mesoscale gap is being bridged from both sides (Goodsell et al., 2020; Liu et al., 2022b) (Figure 2). Indeed, recent live cell MinFlux imaging of kinesin stepping along microtubules highlights the potential of SRM for dynamic analysis of molecular structure (Deguchi et al., 2023).

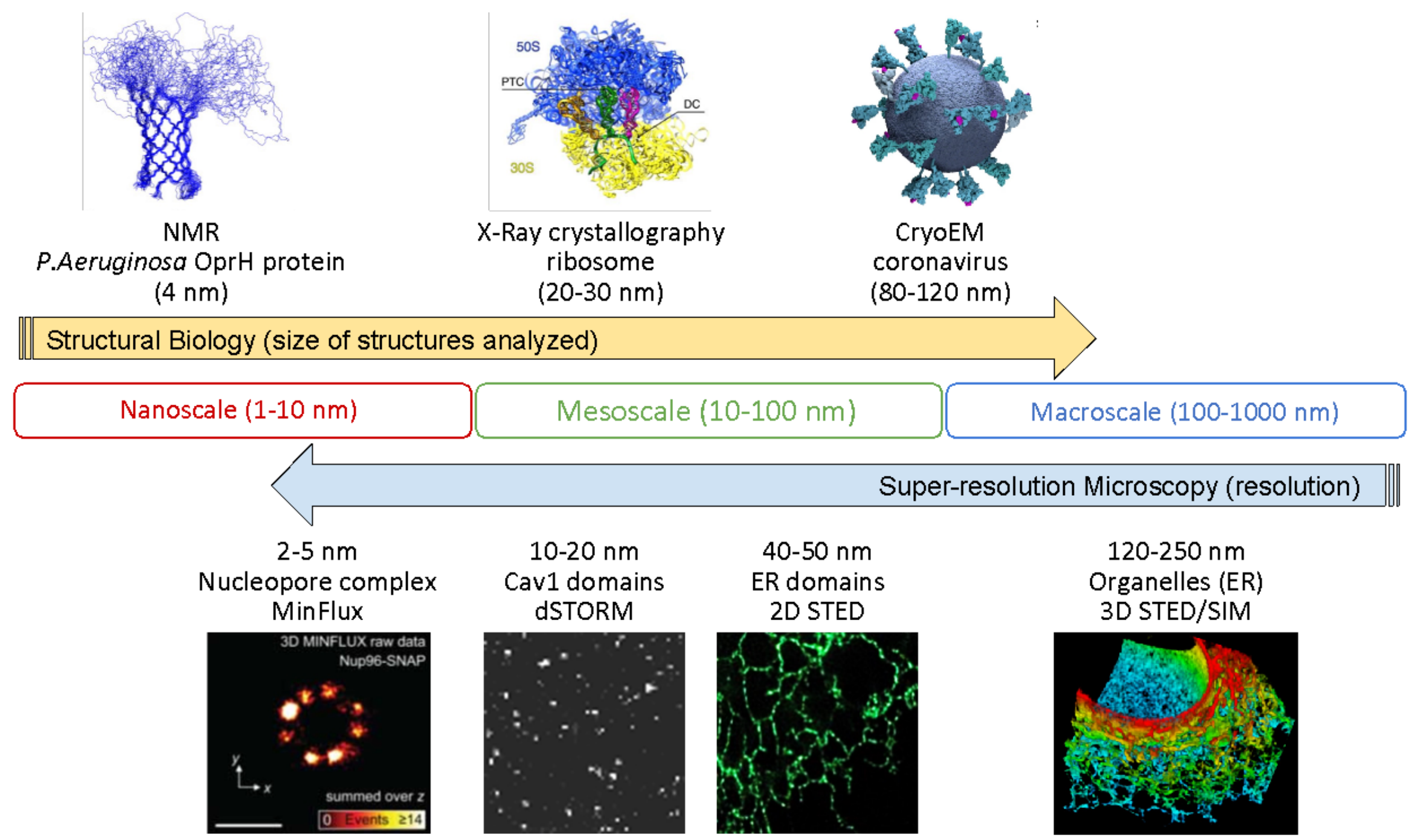

***Figure 2. Super-resolution microscopy and structural biology bridge the mesoscale domain.*** *Structural biology approaches (NMR, X-ray crystallography, cryoEM) with Angstrom level resolution analyze structures as large as the 80-120 nm coronavirus. Whole cell super-resolution microscopy approaches (structured illumination (SIM), stimulated emission depletion (STED), single molecule localization microscopy (SMLM), MinFlux; images (Edrington et al., 2011; Gao et al., 2019; Ke et al., 2020; Khater et al., 2018; Rozov et al., 2019; Schmidt et al., 2021)) enable whole cell analysis with increasing resolution broaching the nanoscale.*

The extensive application of AI to microscopy image acquisition and analysis, potentially leading to intelligent microscopes, has been extensively reviewed (Liu et al., 2021; Morgado et al., 2024; Pylvänäinen et al., 2023). Single molecule reconstruction has been an active area of research and includes application of a convolutional neural network to interpret single molecule localizations directly from SMLM images (Deep-STORM, DBlink) and implementation of anti-bunching to resolve closely spaced emitters, a major issue in single molecule imaging approaches (Kudyshev et al., 2023; Nehme et al., 2018; Saguy et al., 2023; Yang et al., 2021). Recent application of AI to reconstruction methods has the potential to reduce noise, improve spatial and temporal resolution and automate high-throughput super resolution and live imaging (Fu et al., 2023; Priessner et al., 2024; Qiao et al., 2023). Application of AI and machine learning to super-resolution microscopy has been proposed to be a next step in advancing nanomedicine development, drug discovery and antiviral research (Li et al., 2024a; Ortiz-Perez et al., 2024;

Petkidis et al., 2023). In this Perspective, we focus on the specific application of AI to acquire semantic insight, new biology, from super-resolution microscopy.

***Harnessing the power of AI for semantic insight from SRM***

Pixel- and voxel-based SRM approaches, such as structured illumination (SIM) and stimulated emission depletion (STED) microscopy, provide increasingly detailed views of the subcellular space. However, they do not expose the underlying internal construction of subcellular structures. Single molecule localization microscopy (SMLM) approaches, on the other hand, based on analysis of stochastic blinking of isolated fluorophores (Betzig et al., 2006), do not generate images but rather an event list of localizations, or a point cloud. Starting from a point cloud, one can create networks which in turn are well suited to graph-based analysis by machine and deep learning approaches. Despite this well-established network modelling approach, it has been common for many SMLM users to transform the point cloud data generated by SMLM into a pixelated image (Ruszczycki and Bernas, 2018). However, the end user should be aware of the pitfalls in these conversions (Ruszczycki and Bernas, 2018), whether for visualization or quantification.

Indeed, increasing the resolution of pixelated SRM images is akin to providing more enhanced detail, but only of the outer structure of a building (Figure 3). Bridging the mesoscopic gap from diffraction limited fluorescent microscopy to structural biology via SRM therefore requires visualizing inner structure, or molecular architecture. However, in contrast to buildings where the design is generated before construction, architecture of mesoscopic macromolecular structures and organelles is unknown, at best incomplete, and is what biological research endeavours to uncover. Fundamentally, novel biological discoveries lack ground truth (Figure 3).

While SRM may have no definitive ground truth, the actual biology does not exist in isolation. SRM explores spatial and temporal data that, while yet to be described, occurs in the context of an existing accepted body of knowledge obtained by lower resolution confocal microscopy, higher resolution EM, as well as biochemical and structural analyses. SRM datasets, whether point cloud or pixel/voxel-based, generate large data sets including 3D information, time, and multiple channels at nanometer scales. Machine learning identification of patterns and differences from these large datasets can provide novel insight into subcellular structure and molecular architecture.

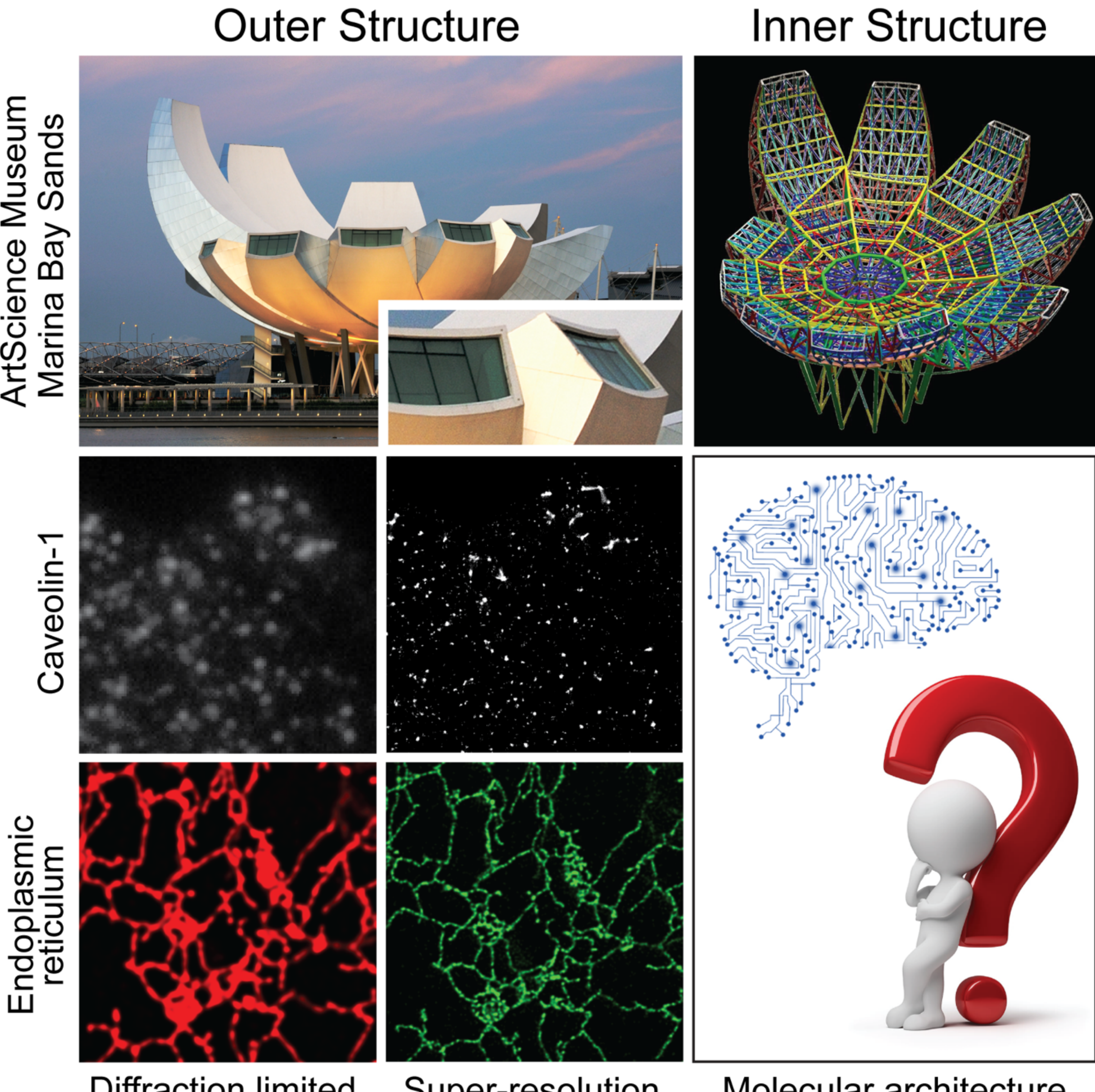

***Figure 3. Molecular architecture by SRM.*** *The top row shows outer (left) and inner (right) structure views of the ArtScience Museum at Marina Bay Sands. Below, pixel-based representation of SRM data provides unprecedented high-resolution views of subcellular structures, such as caveolin-1 domains, (wide-field on left vs SMLM on right) or the endoplasmic reticulum (confocal in red vs STED in green) but remain analogous to enlarged, more detailed views of the external face, or outer structure, of buildings (inset). Ongoing AI-based semantic analysis of SRM will provide the means to explore the design basis, or molecular architecture, of subcellular macromolecules and organelles.*

Machine learning approaches have been exploited to obtain biological insight from both pixel and point cloud super-resolution data. MiLeSIM (Machine Learning Structured Illumination Microscopy) used a supervised machine learning based classifier to extract shape and size features in different strains of live attenuated influenza virus vaccines for high throughput imaging and assessment of viral production (Laine et al., 2018). SRM based image-level class labels of endoplasmic reticulum (ER) were used to train a deep learning model to distinguish between Zika-infected and non-infected cells and showed that discriminating regions correspond to tubular matrix ER morphology (Long et al., 2020). Using a convolutional neural network model, SMLM image stacks are inputted to directly measure molecular diffusion in supported lipid bilayers (Park et al., 2023). Neural network extraction of features from nearest-neighbor distance-derived data identified cluster segmentation in SMLM point cloud space of c-terminal src kinase (CSK) or the associated PAG protein clustering on the cell membrane of T-cells (Williamson et al., 2020). Other advances involve the development of more generally applicable SMLM analysis tools such as SuperResNET and SEMORE (SEgmentation and MORphological fingerprinting), to extract biological features from protein clusters imaged by SMLM (Bender et al., 2024; Khater et al., 2018; Li et al., 2024b; Wong et al., 2024).

Pairing machine learning with other biological information has also been applied to SMLM imaging to develop new insight. LocoMoFit applies a maximum likelihood estimation that fits a provided model to a 3D point cloud structure (Wu et al., 2023). By fitting models, developed based on established structures, to individual coat structures obtained by SMLM, Locomofit supported the idea of a novel cooperative curvature model for clathrin endocytosis (Mund et al., 2023). Another approach used deep learning to pair single molecule imaging of 3D chromatin structure using fluorescent oligo hybridization DNA probes with RNA expression measured by RNA FISH within the same cell, and thereby predicted transcriptional states from the 3D structure of chromatin (Rajpurkar et al., 2021). A workflow developed for 3D reconstruction from 2D SMLM produced a 3D model of the centriole using iterative multi-reference refinement and dual color SMLM imaging between combinations of centriolar proteins allowed for novel insight into protein organization within the centriole (Sieben et al., 2018).

Applying machine learning to SRM requires approaches that do not depend on subcellular object-level ground truth for validation or algorithm training. Indeed, our AI-based image analysis approach for biological discovery and validation has been to avoid pixel-/voxel-/localization-level annotation and use ground truth label annotations of groups of images of cells (e.g. wild-type vs mutant; over vs under-expression, infected vs uninfected cells) (Cardoen et al., 2022; Khater et al., 2018; Long et al., 2020; Saberian et al., 2022) each with prior knowledge of their biological features and functions. AI trained using reliable differential group labels is then used to predict pixel-/voxel-/localization labels. Key to weakly supervised approaches is the trustworthiness of labels, corresponding to phenotypic changes in the images confirmed by other modalities. Prediction of known biological features instills trust in AI for novel biological discovery (Figure 4).

Fundamentally, within this context, weak supervision inspired approaches for SRM could be based on three key principles:

1. Selection of trustworthy group labels grounded in biology;
2. AI-based identification of differential features across groups;
3. Use of *a priori* biological knowledge to corroborate AI-based discovery.

***AI identification of novel biology***

Current weakly supervised approaches are frequently characterized by modularly designed pipelines that, for example, require denoising, segmentation and classification before reporting final results. Development of end-to-end approaches that take in the raw data and report the desired output without explicit intermediate stages can be beneficial when intermediate stages, such as segmentation, are either infeasible or too costly. However, end-to-end methods can require a final `fitting' or calibration stage, where the method's output is rescaled to correspond with validation data. Here we present examples of weakly supervised approaches that incorporate modular design and end-to-end approaches.

*Discovering a novel caveolin-1 scaffold domain*

The protein caveolin-1 (CAV1) is the coat protein for 50-100 nm plasma membrane invaginations called caveolae. Caveolae formation requires the adaptor protein Cavin-1; in the absence of Cavin-1, CAV1 forms functional surface domains called scaffolds (Hill et al., 2008; Lajoie et al., 2009). While caveolae invaginations are morphologically distinct by EM, flat CAV1-positive domains are difficult to detect morphologically by EM and their identification by antibody labeling suffers from the poor antigenicity of most EM approaches. By diffraction limited confocal microscopy, both caveolae and scaffolds present punctate surface labels; differentiating the two by the presence of Cavin-1 assumes Cavin-1 selectively associates with caveolae, which may not be correct (Khater et al., 2019a). Definitive identification of the two structures is problematic. The PC3 prostate cancer cell line, expressing elevated levels of CAV1 in the genetic absence of Cavin-1 and, therefore, of caveolae, together with PC3 cells transfected with Cavin-1 and expressing caveolae (Hill et al., 2008), provide cellular group labels for the identification of caveolae from SMLM data sets by machine learning.

A computational pipeline, SuperResNET (Figure 4), inputs 3D single molecule point cloud data from dSTORM (direct stochastic optical reconstruction microscopy) labeling of CAV1, merges localizations below the resolution limit (20 nm) and filters localizations relative to randomized distributions. Feature-based cluster analysis of interactions below 80 nm identified two groups of blobs in PC3 cells and found two additional groups in Cavin-1 transfected PC3 cells, one of which was significantly larger than the PC3 groups, therefore corresponding to caveolae (Khater et al., 2018). Our validation of AI trained on discriminative group labels (cell lines), in this case, took the form of successfully identified clusters corresponding to the missing component (caveolae) in one of the groups (Khater et al., 2019b; Khater et al., 2018) (Figure 4). Network analysis also identified three non-caveolar CAV1 scaffolds including small S1A scaffolds corresponding to 8S CAV1 complexes whose structure was recently reported by cryoEM (Han et al., 2020; Khater et al., 2018). Caveolae clusters consisted of 12-14 8S CAV1 complexes, matching the dodecahedral structure for caveolae reported by cryoEM (Khater et al., 2019b; Stoeber et al., 2016). The approach itself is therefore validated based on identification of known structures: 8S CAV1 complexes and caveolae.

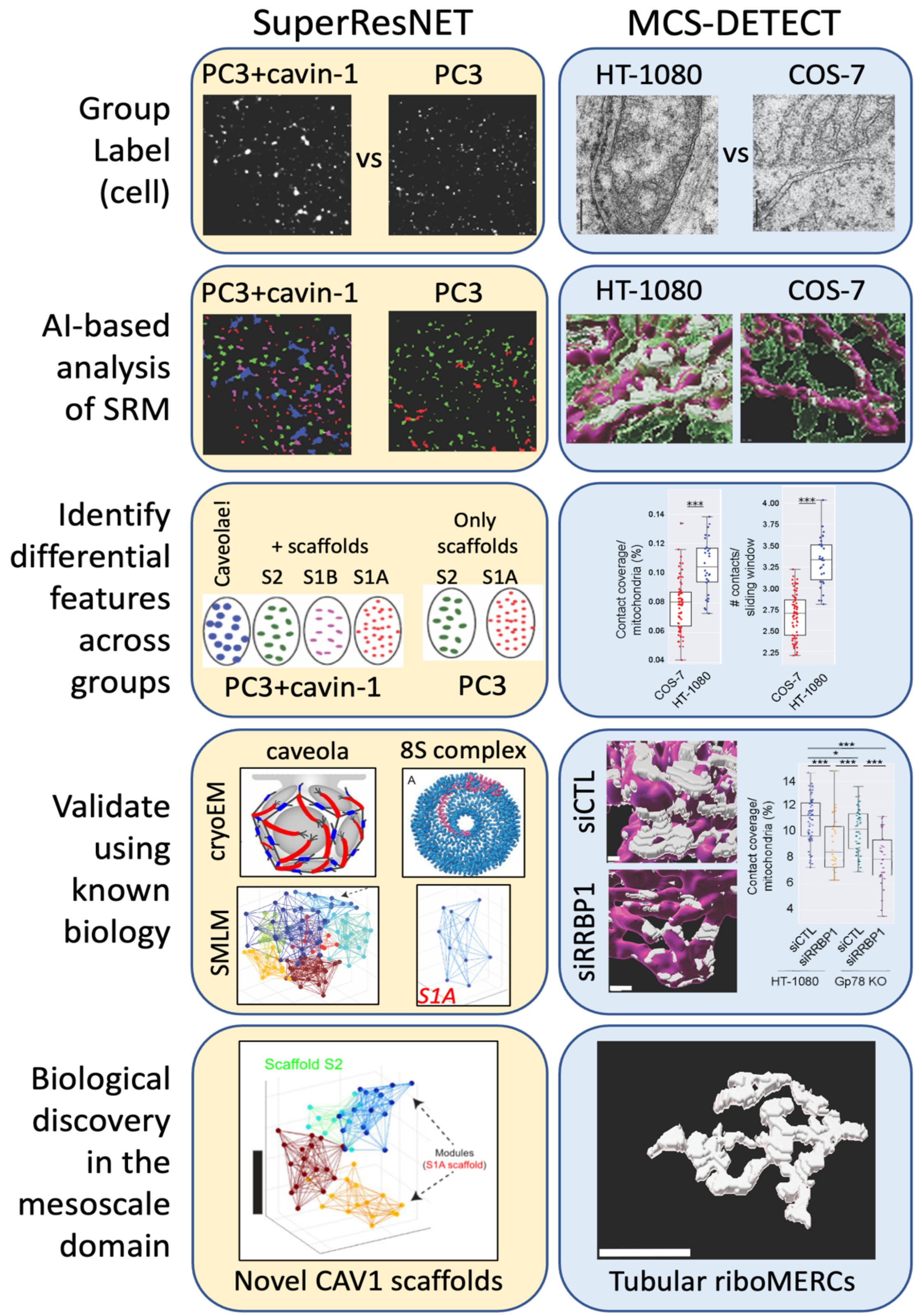

***Figure 4. The weakly supervised paradigm for AI-based semantic analysis of SRM datasets.*** *In the absence of pixel or object level ground truth (strong supervision), prior biological knowledge defines group labels for weakly supervised training of SRM datasets to identify group-specific protein structures. Left: caveolae expression is known to require CAV1 and the adaptor protein cavin-1 (1) (Anderson, 1998; Hill et al., 2008). PC3 cells lack cavin-1 and therefore caveolae; native and Cavin-1 transfected PC3 cells therefore provided group labels for weakly supervised network analysis: SuperResNET (2). Using a proximity threshold that most significantly distinguished point clouds of these two biologically distinct groups, we then segmented clusters of points into blobs (structures or objects), based on 28 features describing size, shape, topology and network measures. Four groups of blobs were found in Cavin-1 expressing PC3 cells, one of which was significantly larger than either of the two groups found in PC3 cells, therefore corresponding to caveolae (3). Structural correspondence of identified structures to known biology, such as 8S complexes and caveolae whose structure has been determined by cryoEM (4) (Han et al., 2020; Stoeber et al., 2016), validates the approach and enables identification of novel structures such as hemi-spherical CAV1 S2 scaffolds (5) (Khater et al., 2019b; Khater et al., 2018). Right: HT-1080 and COS-7 cell lines differ in that only HT-1080 express elongated ribosome-studded mitochondria-ER contact sites (riboMERCS). Based on the differential expression of riboMERCs between these two cell lines, we developed a novel segmentation free algorithm able to reconstruct MERCS, MCS-DETECT, that was optimize by maximizing the differential result between both cell lines (Cardoen et al., 2023b). Feature analysis of MERCs matches differences between the two cell lines based on EM (i.e. ground truth) and the approach is validated by knockdown of a known riboMERC tether, RRBP1 (Hung et al., 2017). MCS-DETECT led to identification of extended tubular riboMERCs.*

Beyond validating known biology (absence of caveolae in PC3 cells), network analysis of single molecule SRM led to the identification of previously unidentified non-caveolar CAV1 scaffolds (Khater et al., 2018). Modularity analysis, in which clusters are broken down into smaller more closely associated clusters, showed that 8S CAV1 complexes combine to form caveolae as well as intermediate scaffold structures. These include previously undescribed 8S CAV1 complex dimers (S1B scaffolds) as well as larger hemispherical S2 scaffolds (Khater et al., 2019b) (Figure 4). Intermediate scaffold structures are supported by the presence of a shoulder on the 8S CAV1 peak in fractionation studies (Hayer et al., 2010), by STED using belief-theory based weakly supervised object detection (Cardoen et al., 2022) and recent identification of CAV1 invaginations in the absence of cavin-1 called dolines (Lolo et al., 2023).

*Identifying sub-pixel resolution riboMERCs*

More recently, we developed a sub-pixel resolution approach to detect membrane contact sites, where the membrane of two organelles approach to within 10-30 nm (Helle et al., 2013). Mitochondria-ER contacts (MERCs) have been well-characterized by EM however their analysis by fluorescent microscopy is challenged by the fact that the distance between the two organelles (10-60 nm) is smaller than diffraction limited resolution and even that of 3D super-resolution approaches such as STED and SIM (Scorrano et al., 2019). To address this, we built on previous work (Cardoen et al., 2022; Cardoen et al., 2020) using Laplacian detection of local intensity changes to detect regions in 3D STED super-resolution images where the intensity of ER and mitochondria change in tandem. Independent of image segmentation, MCS-DETECT sensitively and robustly detects membrane contact sites independently of variations in local signal intensity or background (Figure 4) (Cardoen et al., 2023b).

However, be it diffraction limited colocalization or sub-pixel MCS-DETECT detection of nanometre scale contact sites, validation that contact sites are accurately detected remains challenging. As such, we tuned MCS-DETECT using a group label comparison of HT-1080 cells known to express a distinct type of ribosome-studded MERC (riboMERCs) (Wang et al., 2015) and COS-7 cells that do not express riboMERCs, based on quantitative EM validation (Cardoen et al., 2023b). This weakly supervised approach demonstrated that riboMERC expression is controlled by the RRBP1-SYNJ2BP tether (Hung et al., 2017) and that riboMERC size is regulated by expression of the Gp78 ubiquitin ligase. MCS-DETECT further identified a convoluted tubular morphology for Gp78-dependent riboMERCs (Cardoen et al., 2023b), similar but less extended as the wrapper riboMERCs described in liver by 3D EM tomography (Anastasia et al., 2021). Using group labels based on known biology (i.e. EM), MCS-DETECT has defined a distinct morphology for riboMERCs and identified how their expression and size is controlled. MCS-DETECT, as a probabilistic reconstruction algorithm omitting segmentation, is an example of the move towards end-to-end approaches. However, it still requires specifically designed preprocessing filters to exclude false positives. In addition, the resulting reconstructed membrane contacts are not classified, so MCS-DETECT is not yet a complete end-to-end approach.

MERCs are functionally diverse, involved in ER-mitochondrial calcium exchange and regulation of mitochondrial metabolism, phospholipid and sterol biosynthesis, as well as induction of apoptosis (Rowland and Voeltz, 2012). MERC formation is controlled by numerous tethers (Herrera-Cruz and Simmen, 2017) suggesting that there may very well be multiple sub-classes of MERCs. RiboMERCs represent a sub-class of morphologically-distinct MERCs that provided useful parameters/group labels to define a novel MERC detection approach. Further definition of the other MCS-DETECT detected MERCs may require additional information aside from the relatively limited size and shape features used to characterize riboMERCs (Cardoen et al., 2023b). This would extend to defining MERCs based on their molecular composition, or to live cell analyses to detect dynamics or functional outputs of individual MERCs using fluorescent reporters and high speed SRM (Nixon-Abell et al., 2016). Deep learning could be applied to detect differences in MERC expression patterns due to loss of specific tethers. A critical obstacle, however, is the extreme imbalance in whole cell volumetric data, where larger 3D objects that are discriminative are by definition far less frequent, even though they are visually striking to a human expert. Whether current 3D spatial and temporal resolution limits for voxel-based imaging are sufficient to detect the more subtle distinctions amongst smaller MERCs remains to be determined, and one can only look forward to future hardware, and software, developments that improve our ability to see more clearly within the cell.

Weakly supervised approaches can exploit combinatorial information, that is often `nested’ or `hierarchical’. For example, a single cell can have multiple labels: cell line, expression, treatment, and so on. Furthermore, biological structures are often modular, with each modular part having a label that is hierarchically more refined with respect to its top label. The modular construction of caveolae is a great example of this. Use of weakly supervised learning paradigms to leverage such information, beyond the dual paradigms (i.e. PC3 vs PC3+cavin-1; HT-1080 vs. COS-7) exploited here (Figure 4), will necessarily lead to a more refined description and improved biological understanding of the studied structures.

***Explainable AI in SRM***

If our goal is to determine the inner structure or molecular architecture of biological structures (Figure 3), then it is critical to understand the basis by which AI is making decisions. Raw image data (i.e. pixels, point clouds) are very large; training the model requires preparing the data in a format through which the machine learning algorithm can best identify biologically relevant differences. In the CAV1 point cloud data, 28 features are extracted from the segmented clusters and report on structural aspects of the clusters such as hollowness, topology, network interactions, size, and shape (Khater et al., 2018). This predefined feature approach reports on clearly understood structural aspects of the CAV1 clusters, showing that caveolae are large hollow blobs. Feature analysis of MERCs was limited to three features that could be validated by correspondence to EM (Cardoen et al., 2023b).

Machine learning via handcrafted features is sometimes referred to as shallow learning, since extraction of these features amounts to the execution of a particular known formula or short (shallow) recipe. This contrasts with deep learning, where the exact formula or recipe to extract the features is unknown beforehand; only its general form is known and takes the shape of a long (deep) sequence of operations whose exact equations are optimized to attain a certain objective, the accurate classification of images. Traditionally, deep models are based on convolutional neural networks (CNN) that implement feature extraction primarily via the application of many convolution operations in sequence (layers). The fact that the features are constructed from a deep sequence of operations and determined by a large number of parameters make them hard to interpret and the decision process using these features (i.e. classification) difficult to understand — hence the "black box" label associated with deep models. Over the last decade the number of parameters of deep models has increased by several orders of magnitude, from few layers with millions of parameters, to thousands of layers with trillions of parameters.

Deep learning is achieving state-of-the-art results on a wide array of prediction applications (e.g. classification and segmentation), surpassing their shallow machine learning counterparts often by big margins, and even meeting or surpassing human experts on biomedical image interpretation tasks (Fujisawa et al., 2019; Rajpurkar et al., 2017; Zhang et al., 2019). While deep learning has found rapid adoption in SRM acquisition and image generation related tasks (Hyun and Kim, 2023), discovery oriented SRM data analysis is still limited (Khater et al., 2019a; Zehtabian et al., 2022). Explainable AI (XAI) is a fast-growing field aimed at improving our understanding of deep features and explaining deep model decision processes. XAI is moving towards standardizing the characteristics that an explainable or interpretable model should satisfy (Jin et al., 2023). Beyond natural image analysis, XAI has found rapid adoption in medical image analysis (Chaddad et al., 2023), a field that shares defining characteristics of SRM-based analysis: scarcity of ground truth, targeted at discovery, and high societal impact of findings.

Extending deep learning models of SRM with XAI can enable identification of novel sub-cellular structures and processes (Long et al., 2020; Nagao et al., 2020).

Region based explanations, for example gradient-weighted class activation mapping (Grad-CAM) and its many variants, were some of the earliest XAI approaches finding widespread adoption, as indicated from the citations of the introducing paper (Selvaraju et al., 2020). In contrast, Shapley-based approaches (Shapley, 1953) focus on establishing which features are strongly associated with the outcome of a model (van Zyl et al., 2024). Region and feature based explanations still require expert interpretation and may not always be robust. XAI methods can be difficult to use as evidence or validation for new discoveries in a context where validation is scarce. To ease this burden of uncertainty on expert users, recent advances focus on taxonomies and metrics that differentiate types of XAI (Ibrahim and Shafiq, 2023). More recently, XAI approaches are exploring large language models (LLM) based higher level explanations, such as providing language based `concepts' (Maddigan et al., 2024; Mavrepis et al., 2024) alongside model performance. Here, the model would explain, using plain English, what concepts or domain specific keywords are associated with the model's decision. It should be noted that using XAI is not necessarily guaranteed to augment AI-guided expert decisions (Jin et al., 2024). Finally, in discovery-oriented research it is important to select methods that are causative, and not just correlative (Vasudevan et al., 2021), aligning with current experiment design where controlled experimental factors such as knockout can ensure elimination of correlative effects in favor of the true causal information.

***Beyond weak supervision***

Beyond weakly supervised methods, self-supervision has emerged as a powerful paradigm to learn rich semantic features that can be specialized for a chosen end-task with a minimum of ground truth labels, and has been successfully adopted to microscopy (Kobayashi et al., 2022; Wu et al., 2022). Unlike weak supervision, self-supervision does not require a group (e.g. cell) level label to learn an informative representation. However, it does not provide an end-task capable solution, as the final stage would be either unsupervised, supervised, or weakly supervised. The advantages, however, include requiring far fewer labels, higher performance, and the ability to reuse the learned encoding for multiple tasks.

Counterfactual learning simulates from existing data 'what if' scenarios, causal relations between objects or conditions without the specific experiment taking place (Pearl, 2010). These then extend to generative models that learn to synthesize data from images, features, or even based on descriptions (language). While generative models do not specifically enforce a causal relation between their input and what they generate, the potential of such models in subcellular biology is largely untapped, but highly promising based on its interpretability and rapid adoption in medical imaging (Yi et al., 2019). These represent powerful paradigms for both medical and biology experiments, potentially enabling experiments otherwise not feasible due to ethical or resource constraints. Scalability is an issue, with SRM data often being orders of magnitude larger in dimension compared to datasets on which most deep learning models are being developed, whether it is for point cloud- or voxel-based data.

A recent perspective (Volpe et al., 2023) argues that stability (i.e. robustness to confounding factors as well as reliability and performance on unseen datasets) of AI models is critical for adoption of unsupervised learning for reconstruction of SRM data. Stability of AI models is critical and can be addressed by the emerging work in continual (Parisi et al., 2019) and out-of-distribution learning, as well as by resolving `short-cuts' that degrade performance on new datasets (Robinson et al., 2021). We would argue that ongoing improvements in both SRM and AI offer tremendous potential not only for SRM data interpretation but also for semantic discovery leading to the accelerated exploration of nanoscale and mesoscale biology.

The potential of foundation and large language models (FLLM) offers great potential as these models are built on enormous datasets, then tuned to domain specific tasks, as recently demonstrated in single cell genomics (Luecken et al., 2022; Myers et al., 2024). However, the challenge in validation, interpretation, and reproduction of such models remains. Problematic here is the rapid increase in compute power, and associated carbon footprint, in using, let alone training, such models (Bouza et al., 2023; Shah and Bhavsar, 2022). Validation and development of any novel technique is still constrained by access to representative, diverse open data that the community can review and challenge. In addition, translating results obtained on benchmark data to newly acquired data is sensitive to acquisition specific signatures, such as PSF configuration, localization algorithm, and specimen specific factors. For example, given that most SRM acquisitions are developed by closed source vendors, it is often challenging to translate results across vendors. Recent efforts to standardize file formats and experimental design have been driven by the scientific community rather than commercial needs and are a critical first step in resolving these challenges (Cardoen et al., 2023a; Sarkans et al., 2021; Schmied et al., 2024). Community driven efforts are underway to provide a single standardized format for critical metadata (Moore et al., 2023; Swedlow et al., 2021). Ignoring such confounding factors will only delay advances in AI driven discovery and limit reproducibility. Open datasets, acquired and curated at great cost in expense, training, and expertise by publicly funded scientific centers, should be standardized and shared with a clear license to favor open science.

Be it determined by human endeavour or AI, newly discovered scientific knowledge needs to be understandable, reproducible, and trustworthy to be adopted by experts in the field. These challenges have been encountered by scientists introducing technological developments, going back to Golgi. The exponential changes that computer science and AI have brought to science, not to mention our everyday lives, means that validation will need to keep pace. Waiting 50 years for validation of scientific discovery, as for the Golgi apparatus, is not an option.

**Acknowledgements**

Collaborative work in the Nabi and Hamarneh labs is supported by grants from the Canadian Institutes for Health Research (AWD-022443; PJT-175112), National Sciences and Engineering Research Council of Canada (RGPIN-2019-05179; RGPIN-2020-06752), and CIHR/NSERC Collaborative Health Research Projects (CPG – 163989; CHRP 538851-19). We thank Ali Bashashati, Gal Harari, Josef Penninger and Calvin Yip for critical review of the text, Y. Lydia Li for the PC3 images, and Moshe Safdie for the images of the ArtScience Museum at Marina Bay Sands and insightful discussions. Images were adapted from: 10.1074/jbc.M111.280933; 10.1038/s41467-019-10409-4; 10.1038/s41586-020-2665-2; 10.1038/s41598-019-46174-z; 10.1038/s41467-021-21652-z.